\documentclass[11pt]{article}

\usepackage{epsfig}
\usepackage{graphicx}
\usepackage{cancel}
\usepackage{amssymb,amsmath}

\newcommand{\be}{\begin{equation}}
\newcommand{\ee}{\end{equation}}
\newcommand{\bea}{\begin{eqnarray}}
\newcommand{\eea}{\end{eqnarray}}

\newcommand{\II}{\mathbb{I}}

\newcommand{\s}{\mathbf{s}}
\newcommand{\Sc}{\mathcal{S}}

\newcommand{\Dl}{\Delta_l }
\newcommand{\Dr}{\Delta_r }
\newcommand{\DV}{\Delta V}
\newcommand{\DL}{L}

\newcommand{\EDL}{\overline{\DL}}
\newcommand{\EU}{\overline{U}}
\newcommand{\EA}{\overline{A}}
\newcommand{\E}[1]{\mathbb{E}\left[ #1 \right] }
\newcommand{\e}{\varepsilon}
\newcommand{\CM}{\mbox{CM}}
\newcommand{\eref}[1]{(\ref{#1})}
\newcommand{\hp}{\hat{p}}

\begin{document}
\begin{titlepage}
\begin{flushright}
\end{flushright}
\vskip1cm
\begin{center}
{\Huge
Central Counterparty Risk}
\vskip1.0cm
{\Large Matthias Arnsdorf }
\vskip0.5cm
Quantitative Research, JP Morgan
\vskip0.1cm
Email: matthias.x.arnsdorf@jpmorgan.com
\vskip0.1cm
November 2011 \\

\vskip1.0cm
{\Large Abstract:\\}
\end{center}
\parbox[t]{\textwidth}{

A clearing member of a Central Counterparty (CCP) is exposed to losses on their default fund and initial margin contributions. Such losses can be incurred whenever the CCP has insufficient funds to unwind the portfolio of a defaulting clearing member. This does not necessarily require the default of the CCP itself.
In this note we aim to quantify the risk a financial institution has when facing a CCP.

We show that a clearing member's CCP risk is given by a sum of exposures to each of the other clearing members.
This arises because of the implicit default insurance that each member has provided in the form of mutualised, loss sharing collateral.
We calculate the exposures by explicitly modeling the capital structure of a CCP as well as the loss distributions of the individual member portfolios.

An important consideration in designing the model is the limited transparency with respect to the portfolio composition and collateral levels of individual clearing members. To overcome this we leverage the fact that, for a typical CCP, margin levels are risk-based. In particular, we parameterise the  portfolio loss tail as a Pareto distribution and
we calibrate this to the CCP defined probability of losses exceeding the posted initial margin levels.

A key aspect of the model is that we explicitly
take into account wrong-way risk, i.e.\ the fact that member defaults are more likely to occur in stressed market conditions, as well as potential contagion between a member's default and the losses on his portfolio.
}

\vspace{1.5cm}
\newcounter{helpfootnote}
\setcounter{helpfootnote}{\thefootnote}
\renewcommand{\thefootnote}{\fnsymbol{footnote}}
\setcounter{footnote}{0} \footnotetext{ Opinions expressed in this
paper are those of the author, and do not  necessarily reflect the
view of JP Morgan. We would like to thank Andrew Abrahams for originating the project on CCP risk and for shaping many of the key ideas. We also thank
 Regis Guichard, Joe Holderness, Mitchell Smith, Marnie Rosenberg and Rajalakshmi Ramanath for valuable discussions.

}
\renewcommand{\thefootnote}{\arabic{footnote}}
\setcounter{footnote}{\thehelpfootnote}

 \end{titlepage}


\section{Introduction}
The financial crisis of 2008 has resulted in a concerted regulatory drive to substantially increase the proportion of derivatives that are centrally cleared. The primary motivation is to reduce bilateral counterparty risk, increase transparency and avoid contagion in the case of the default of a large financial institution.

Central counterparties (CCPs) are designed to mitigate counterparty risk through multi-lateral netting, high levels of collateralisation as well as loss mutualisation (c.f.~\cite{Duffie2010} for a quantitative comparison of the efficiency of multi-lateral vs.\ bi-lateral netting). Indeed, the default of a CCP is a rare event and only three have failed in recent times: the Caisse de Liquidation Paris in 1974,
the Kuala Lumpur Commodity Clearing House in 1983 and the Hong Kong Futures Exchange due to the 1987 market crash (c.f.~\cite{Tucker2011}).

However, this does not mean that a clearing house is riskless. A CCP is a risk sharing arrangement and each member is liable for the performance of all the other members. In moving from bilateral to central clearing, the risk of counterparty default has been transformed, to a large extent, into the risk of losses on the mutualised collateral pool (default fund) that each member has contributed to in order to protect the CCP from default. In particular, in the event of a clearing member default, all uncollateralised losses arising from the liquidation of the member's portfolio will be shared pro-rata amongst the surviving members.

Given the increasing volume and complexity of cleared trades as well as the systemic role of CCPs, it is important that the risks a clearing member faces are understood and quantified.
Even if loss rates and probabilities should be low due to high levels of collateralisation, the absolute exposure amounts are likely to be very significant. Past crises have shown that it is dangerous and costly to underestimate tail risks.

An important difference between bilateral counterparty risk and the risk facing a CCP is that the CCP risk is not primarily driven by the exposure on a member's own portfolio. The member's CCP risk can increase even if his portfolio does not change.
This is because each member has  provided insurance on the tail losses of all other clearing members. Indeed, we will show that a member's exposure can be written as a sum of exposures to each clearing member.

These exposures are naturally hedged by buying CDS protection on each of the members. The amount of protection required depends on the margin and collateral levels held by the CCP as well as the size of the tails of the portfolio loss distributions, which are specific to the market in which the CCP is operating. The total cost of protection gives the value of the CCP risk.

This means that for a clearing member to asses his risk towards a CCP he needs information about the portfolio composition of all other clearing members as well as their posted collateral levels. This information is typically confidential. What is available to a member are the details of his own portfolio, total margin levels held by the CCP as well as identities of the other clearing members. The aim of this paper is to provide a practical model that can give realistic risk estimates based only on the limited amount of available public data.

We do this by leveraging the risk management framework of the CCP. In particular, we make use of the fact that it is now standard practice to set initial margin levels using a value-at-risk (VaR) model. This means that, for a given portfolio, the probability of losses exceeding the posted initial margin levels over the liquidation period is given by a fixed probability. This is set by the CCP and is typically around 1\%.
This information can be used to fit a parameterised distribution to the tail of the portfolio losses. Here we choose a Pareto distribution to capture the heavy tails exhibited by the time series of financial returns.

An important feature of our model is that we explicitly consider wrong-way risk and contagion. Wrong-way risk arises due to the fact that clearing member defaults are more likely in times of stress and hence when portfolio losses are large. Contagion refers to the possibility that the member default itself will cause a shock to the market. This means that losses can be larger than  is implied by the confidence level used in setting margin levels. These two effects are key risk factors driving the expected losses due to CCP membership (c.f.~\cite{Bates1999} for an analysis of CCP expected losses before and after the 1987 market crash).

\section{Central Counterparty Structure and Risk}
We begin our discussion by taking a closer look at how a clearing house is designed and how this relates to the risks that a clearing member has when dealing with a CCP. A good overview can be found in~\cite{Knott2002}.

\subsection{Risk Waterfall}
A typical CCP has a multi-layer capital structure to protect itself and its members from losses  due to member defaults. In general, the following types of collateral will be held:
\begin{description}
\item[Variation Margin:] Variation margin is charged or credited daily to clearing member accounts to cover any portfolio mark-to-market (MtM) changes.
\item[Initial Margin:] Initial margin is posted by clearing members to the CCP. This is to cover any losses incurred in the unwinding of a defaulting member's portfolio. Typically the margin is set to cover all losses up to a pre-defined confidence level in normal market conditions.
\item[CCP Equity:] A typical CCP will have an equity buffer provided by shareholders. The position of the equity buffer in the capital structure can vary between CCPs.
\item[Default Fund (funded):] Every member contributes to the clearing house default fund. This acts as a form of mutualised insurance for uncollateralised losses.
\item[Default Fund (unfunded):] In addition to the default fund contributions that have been posted to the CCP, each clearing member is usually committed to providing further funds if necessary. The maximum amount of additional funds that can be called upon depends on the CCP. In some cases the liability is uncapped.
\end{description}

Losses arising from a member default will first be covered by the defaulting member's initial margin and default fund contribution. Uncollateralised losses will then be charged against the CCP's equity and ultimately the mutualised default fund.
If all funds are used up and there are still outstanding losses then the CCP could find itself in default.

\subsection{Default fund risk}
If, following a clearing member default, there are losses on the mutualised part of the default fund then all surviving clearing members are required to recapitalise the fund under CCP rules. Losses are allocated pro-rata to the surviving members. For example, this can be on the basis of a member's share of the total default fund pool, total notional or total risk. However, it does not depend on the performance of the surviving members' portfolios at the time of default.

There can be a lag between the clearing member default and any potential recapitalisation of the default fund. We refer to this as the recapitalisation or allocation period. Depending on the CCP, this can range from 0 to ca.\ 30 days (a period of 0 means that losses need to be recapitalised for each default). At the end of the period the default fund will be recapitalised for losses due to all defaults in the period.

A financial institution can incur losses due to clearing member defaults as long as they are members of the CCP.
Since the default fund is recapitalised after each recapitalisation period, it follows that the default fund contribution at risk does not necessarily reduce post a default event.

\subsection{CCP default risk}
If losses due to member defaults exceed all funded and unfunded default fund contributions, then the CCP can be in default itself. Given the high levels of collateralisation that a CCP can draw upon, this is an extreme tail event.

If a CCP default should occur, the CCP assets will be used to cover the liabilities the CCP has to the clearing members.
Assets potentially available to the CCP in default are the posted initial margin contributions that are not held in a segregated, default-remote account.

If all of the initial margin is ringfenced then a surviving clearing member will be able to recover the full margin posted. However, any positive P\&L on the clearing member's portfolio will be lost.

If the margin is not ringfenced then the pooled margin contributions will be used to offset any MtM losses on the clearing member's portfolio. However, the surviving members stand to lose a portion of their posted initial margin. Total CCP losses would be allocated to each member pro-rata based on e.g. the share of the posted initial margin, default fund contribution or total notional.  As with the default fund allocation this is independent of the value of the clearing member's portfolio at the time of allocation. In particular, a surviving member can incur a loss even if his portfolio MtM is negative.

In the following we assume that the initial margin is \emph{not} ringfenced. Even if legally this might be the case it is unclear that this is enforceable in practice. In addition, typically only a portion the margin is ringfenced.

\section{CCP Loss Model}
In this section we develop our model of a clearing member's CCP risk.

We consider a central counterparty with $N+1$ clearing members $\CM_k$ with $k \in \{0,..,N\}$.
At each point in time, $t$, the posted initial margin of member $k$ is denoted by $M_k(t)$. The funded default fund contribution is denoted by $D_k(t)$. The total default fund size is thus  given by $D_{tot}(t) \equiv \sum_{k=0}^{N} D_k(t)$. The CCP equity cushion will be denoted by $E$.

We now take the position of
clearing member $\CM_0$. Our aim is to calculate the cost of the risk, $C_0(T)$, that member $\CM_0$ has up to a horizon $T$ due to membership of the CCP. This will be defined as the member's discounted expected loss up to time $T$.

\subsection{Portfolio losses}
Let us assume that clearing member $k$ has defaulted at time $t = \tau_k$. We denote the value of the portfolio of member $k$ at time $t$ by $V_k(t)$. Post-default the portfolio will be unwound over the liquidation period $\Dl$. This is typically of the order of 2 to 5 days given that cleared products tend to be very liquid.  Note that up to time $\tau_k$ the portfolio value $V_k(t)$ will have been collateralised by the variation margin. This means that the loss on the portfolio post-default is given by the change in value over the liquidation period, i.e.\ by:
\be
 \DV_k(\tau_k) = V_k(\tau_k+\Dl) - V_k(\tau_k).
 \ee
Here and in the following we use the convention that a positive value of $\DV_k$ indicates a loss to the portfolio.

The losses are collateralised by the defaulting members initial margin $M_k(\tau_k)$ and default fund contribution $D_k(\tau_k)$ at the time of default. The uncollateralised loss
\[
U_k(\tau_k) \equiv (\DV_k(\tau_k) - M_k(\tau_k) - D_k(\tau_k))^+
 \]
 will be covered in the first instance by any available CCP equity cushion. All remaining excess losses will be
allocated amongst the surviving clearing members. Note that, in general, $M_k(t)$ and $D_k(t)$ are random variables. The collateral levels will change as a function of market volatility and portfolio composition.

If there is a default at time $t$, then all losses incurred during the recapitalisation period $[t,t+\Dr]$ will be allocated at time $t+\Dr$. Typically $\Dr$ is of the order of 30 days and $\Dr > \Dl$.
 To model this we introduce a timeline, $\{t_0,t_1,...,t_n\}$, up to the horizon $T$ with step $\Dr$, i.e. $t_n = T$ and $t_i = t_0 + i\Dr$. We assume that loss allocations happen only at the timeline points $t_i$.

Let $\s$ denote the set of indices of all members that have defaulted in the period $[t_{i-1}, t_i]$., i.e. for each $j \in \s$ we have $t_{i-1} \leq \tau_j < t_i$. Since we are taking the viewpoint of member $\CM_0$, we are assuming throughout that $\CM_0$ does not default, i.e. $0 \not \in \s$. The total excess loss over the recapitalisation period in the scenario $\s$ is given by:
\be
\DL_{tot}(t_i; \s) \equiv \left(\sum_{j \in \s} U_j(\tau_j) - E\right)^+
\ee
This loss will be allocated amongst the surviving members at time $t_i$. In practice, the mutualised default fund would be used first to cover the losses.
Once both funded and unfunded contributions are exhausted the CCP would find itself in default. As discussed above, we assume that in this case the surviving members initial margin would be used to cover any remaining losses.

In addition, we make the conservative and simplifying assumption that \emph{all} uncollateralised losses will ultimately need to be covered by the surviving members, i.e. we do not cap our loss at our total collateral contribution.
The approximation is unlikely to be material given that the probability of losses exceeding the entire default fund as well as all posted initial margin is very remote. Indeed, we expect the bulk of our losses to be on the default fund.

\subsection{Loss allocation}
We need to specify how the loss, $\DL_{tot}$, is apportioned to the surviving members. A typical convention, which we also follow here, is that the allocation is proportional to our contribution to the default fund as a fraction of the total remaining fund.
 In particular, the fraction of losses allocated to member $\CM_0$ at time $t_i$ is given by:
\be
A_0(t_{i};\s) = \frac{D_{0}(t_{i})}{D_{tot}(t_{i}) - \sum_{j \in \s} D_j(t_{i})}
\ee
and the loss that is incurred in scenario $\s$ over the recapitalisation period is given by:
\be
\DL_0(t_i; \s) = A_0( t_{i}; \s)\DL_{tot}(t_i; \s)
\ee

It is straightforward to adapt our methodology to different allocation choices, e.g. based on initial margin or total collateral. The main point is that the allocated portion increases with the number of defaults.

\subsection{The CCP Counterparty Charge}\label{sec-expLoss}
 We are interested in calculating the expected loss
 \be
 \EDL_0(t_i) \equiv \E{\DL_0(t_i; \s)}
 \ee
 over the period $[t_{i-1},t_i]$.
This expectation is both over the scenarios $\s$ as well as the loss distribution of the member portfolios.

 To formalise this we introduce the total set of default scenarios, $\Sc$, as the set of all unique, non-empty, subsets of $I(N) \equiv \{1,2,...,N\}$ (Note that $\CM_0$ is excluded).
This allows writing the expectation above as:
\be \label{eq-ExpLossPeriod}
\EDL_0(t_i) = \sum_{\s \in \Sc} P(\s;t_i) \E{A_0( t_{i}; \s)\middle| \s} \E{ \DL_{tot}(t_i; \s) \middle| \s}
\ee
where $P(\s; t_i)$ is the probability of the scenario $\s$ occurring in the period $[t_{i-1}, t_i]$. As detailed in the next section, we assume that conditional on the scenario $\s$, the allocation is fixed and independent of the portfolio losses.

To obtain the total expected loss up to horizon $T$ we need to combine the expected losses over the different periods.
After each allocation period the default fund will be recapitalised by the surviving members.
We make the  steady-state assumption that after recapitalisation the risk profile of the CCP is the same as it was before in the sense that defaulting members will have been replaced and their portfolios will have been taken over. In particular, we assume that the expected losses in the different allocation periods are independent.

Note that if the CCP were to default, then our steady-state assumption would clearly no longer hold since in this case our exposure to the CCP would cease. However, the probability of this event is small and hence the approximation is unlikely to be material.

With these assumptions, the CCP risk can be written as the sum over the discounted loss expectations at all timeline points, i.e.:
\be
C_0(T) = \sum_{i=1}^n Z(t_0, t_i) \EDL_0(t_i)
\ee
where $Z(t_0, t_i)$ is the discount factor up to $t_i$, which we have assumed to be independent of the allocated loss.

In the following sections we look at how the expected loss and allocation in equation~\eref{eq-ExpLossPeriod} can be calculated.

\subsection{Expected Collateral Levels}\label{sec-margins}
Margin and default fund levels can vary over time due to changing market conditions as well as changing portfolio composition.

We assume here that the portfolio composition stays constant over time.
 The effect that we need to capture is that, conditional on member default, we expect margin levels to be high. This is because member defaults are more likely to occur in periods of market stress rather than completely idiosyncratically.  This is a wrong-way risk effect familiar from credit value adjustment (CVA) calculations.

 Here we make a simple assumption; namely that in a scenario where clearing member $\CM_k$ has defaulted, the initial margin as well as the default fund contributions are given by stressed levels $M_k^*$ and $D_k^*$, which are related to today's margin levels by a scaling factor. In other words we assume:
 \be
 M_k^* = w M_K(t_0) \equiv w M_k
 \ee
\be
D_k^* = w D_k(t_0) \equiv w D_k
\ee
where $w$ is the \emph{wrong-way factor}  which gives the ratio of the stressed margin over today's margin level. Note that we have assumed for simplicity that the wrong-way factor is constant across time and members and is the same for the initial margin and default fund contribution.
Further motivation for this assumption and details on how $w$ can be estimated are given in section~\ref{sec-w}.

Given these assumptions, the allocation factor only depends on today's default fund levels, i.e.:
\be
\EA_0(\s)  \equiv \E{A_0( t_{i}; \s)\middle| \s} = \frac{D_{0}^*}{D_{tot}^* - \sum_{j \in \s} D_j^*}
=\frac{D_{0}}{D_{tot} - \sum_{j \in \s} D_j}
\ee
where $D_{tot}^* \equiv \sum_{j=0}^N D_j^*$.

 \subsection{Portfolio Expected Loss}\label{sec-portfolioLoss}
 The CCP risk depends on the portfolio loss distributions of all clearing members. In particular, we need to evaluate the expectation:
 \be
 \E{ \DL_{tot}(t_i; \s) \middle| \s} = \E{ \left(\sum_{j \in \s} U_j(\tau_j) - E\right)^+ \middle| \s}
 \ee
 which is conditional on the default scenario $\s$.
 To calculate this we need to know the distribution of $\sum_{j \in \s} U_j(\tau_j)$ which, in turn, depends on the joint loss distribution of all the member portfolios. To simplify the calculation we make the conservative assumption that $E= 0$. An alternative, equally tractable approximation, would be to allocate the equity cushion to the defaulted clearing member's initial margin.

 This means that the expectation is linearised and we have:
 \bea
 \E{ \DL_{tot}(t_i; \s) \middle| \s} &=&  \sum_{j \in \s} \E{U_j(\tau_j) \middle| \s} \\
 &=& \sum_{j \in \s} \E{ (\DV_j(\tau_j) - M_j(\tau_j) - D_j(\tau_j))^+ \middle| \s} \\
 &\equiv& \sum_{j \in \s} \EU_j(\tau_j; \s)
 \eea

Hence, we have reduced the problem to estimating the marginal distributions of the portfolio losses, $\DV_k(t_i)$, \emph{conditional on default}.  Note that using the assumptions of section~\ref{sec-margins} we can write $M_j(\tau_j)$ and $D_j(\tau_j)$ as $M_j^*$ and $D_j^*$ respectively.

We propose to model the tail of the loss distribution as a Pareto distribution.
To be more precise, we assume that for $k \in \s$ and $x \geq M_k^*$ we have:
\be
P[\DV_k(t_i) > x | \s ] = \hat{p}_k(\s, t_i)\left( \frac{M_k^*}{x} \right)^\alpha
\ee
where $\alpha$ is the Pareto index that determines how fast the tail of the distribution decays. This parameter needs to be calibrated to the underlying market as is discussed in section~\ref{sec-alpha}.
The probability of losses exceeding the initial margin in the scenario is given by:
\be
\hp_k(\s, t_i) = P[\DV_k(t_i) > M_k^* | \s ]
\ee
We assume in the following that $\hp_k(\s, t_i) $ does not depend explicitly on time. In addition, we assume that $\hp_k(\s, t_i)$ does not depend on how many or on which clearing members have defaulted, only on the fact that there has been at least one default. This means that we can drop the $\s$ and the $t$ labels and write $\hp_k(\s, t_i) \equiv \hp_k$. Further details on how this input is determined are provided in section~\ref{sec-p}.

It follows that the conditional expected loss, $\EU_k(t_i;s) \equiv \EU_k$, also has no explicit time or scenario dependence.
It is straightforward to calculate $\EU_k$ as:
\bea
\EU_k &\equiv& \E{ (\DV_k - M_k^* - D_k^*)^+ \middle| \s} \\
&=& \hp_k\int_{M_k^*+D_k^*}^{\infty} (x - M_k^* - D_k^*)\alpha M_k^{*\alpha} x^{-(\alpha+1)} dx \\
&=& \frac{\hp_k}{\alpha-1} \left( \frac{M_k^*}{M_k^* + D_k^*} \right)^\alpha (M_k^*+D_k^*) \\
&=& \frac{w \hp_k}{\alpha-1} \left( \frac{M_k}{M_k + D_k} \right)^\alpha (M_k+D_k)
\eea
which is well-defined for $\alpha > 1$.

\subsection{Summing over Scenarios and the Exposure to Clearing Members}

We will now write equation~\eref{eq-ExpLossPeriod} as a sum over exposures to the individual clearing members:
\bea
\EDL_0(t_i) &=& \sum_{\s \in \Sc} P(\s;t_i) \EA_0(\s) \sum_{k\in\s} \EU_k \\
&=& \sum_{\s \in \Sc} P(\s;t_i) \EA_0(\s) \sum_{j = 1}^N \II_{k\in\s} \EU_k \\
&=& \sum_{k = 1}^N \EU_k \sum_{\s \in \Sc} \II_{k\in\s}  P(\s;t_i) \EA_0(\s)   
\eea
where we have introduced the indicator function $\II_x$ which is $1$ if $x$ is true and 0 otherwise.

Note that the sum, $\sum_{\s \in \Sc} \II_{k\in \s}P(\s;t_i)$, is the marginal probability that the member $\CM_k$ defaults in the period $[t_{i-1}, t_i]$ and is given by:
\be\label{eq-MarginalProb}
\sum_{\s \in \Sc} \II_{k\in \s}P(\s;t_i) = \lambda_k(t_i) \Dr
\ee
where $\lambda_k(t_i)$ is the default intensity of the member, which we have assumed constant over $[t_{i-1},t_i]$.
The intensity is given by:
 \be
 \lambda_k(t) = \frac{1}{1-P_k(t)}\frac{dP_k(t)}{dt}
 \ee
 where $P_k(t) = P[\tau_k < t]$ is the marginal default probability of member $\CM_k$. Note that in equation~\eref{eq-MarginalProb} we have implicitly assumed that $\CM_k$ has not defaulted prior to $t_{i-1}$, since we have not multiplied the default intensity with the survival probability up to $t_i$. This is to be understood in the context of our steady state assumption: in the case of default the member would be replaced by an equivalent member.

If $\EA_0(\s)$ did not depend on $\s$ then the expected loss would just depend on the marginal default probabilities of each member.
In order to isolate the scenario dependence of $\EA_0(\s)$
we write for $\s \in \Sc$ and for any $k\in \s$:
\bea
\EA_0(\s) &=& \frac{D_0}{D_{tot} - \sum_{j \in \s} D_j} \\
&=& \frac{D_0}{D_{tot} - D_k}(1+ B_k(\s))
\eea
where:
\be
B_k(\s) \equiv \frac{  \sum_{j \in \s \setminus \{k\} } D_j}{D_{tot} - \sum_{j \in \s } D_j}
\ee
Note that $B_k(\s)$ is 0 unless there are at least two member defaults in the scenario $\s$.

We can now express the expected loss as:
\be \label{eq-LossToMembers}
\EDL_0(t_i) = D_0 \sum_{k =1}^N \frac{\EU_{k}}{D_{tot} - D_k}\left(1 + \e_k(t_i)\right) \lambda_k(t) \Dr
\ee
where the correction term:
\be
\e_k(t) =  \frac{1}{\lambda_k(t) \Dr}\sum_{\s \in \Sc}   P(\s;t)\II_{k\in \s} B_k(\s)
\ee
is a probability weighted sum over scenarios with more than one default.
It corrects for the fact that the loss allocation is higher in multiple default scenarios and depends on the default correlation as well as the size of the allocation period, $\Dr$. It can be calculated given a model for joint member defaults.

We show the dependency of $\e_k$ on the default correlation in table~\ref{table-error}.
This is based on a clearing house with 15 members. All default fund contributions are assumed equal. The member CDS spreads are taken to be 200 bps with a recovery rate of $40\%$. Finally, we use a Gaussian copula to generate joint default probabilities.
We see that $\e_k$ increases exponentially with correlation. The sensitivity to the spread and allocation period is comparatively low.
For a correlation level of 60\% we see a correction term of around 15\% to 20\%.

\begin{table}[h]
\begin{center}
\begin{tabular}{|c||c|c|c|} \hline
$\Dr$ (days)  & 30    & 10   & 30       \\ \hline
Spread (bps) & 200     &  200     & 100       \\ \hline \hline
Correlation & \multicolumn{3}{c|}{$\e_k$} \\ \hline
0\%    & 0\%     & 0\%    & 0\%         \\ \hline
20\%    & 1\%     & 1\%    & 1\%         \\ \hline
40\%    & 5\%     & 3\%    & 4\%         \\ \hline
60\%    & 19\%     & 13\%    & 15\%         \\ \hline
70\%    & 38\%     & 27\%    & 31\%         \\ \hline
80\%    & 81\%     & 62\%    & 69\%         \\ \hline
90\%    & 190\%     & 160\%    & 170\%         \\ \hline
\end{tabular}
\end{center}
\caption{Sensitivity of the correction term, $\e_k$, to default correlation under different spread and allocation period assumptions.} \label{table-error}
\end{table}

\subsection{Summary}

The cost of the risk of CCP membership is given  by:
\be
C_0(T) = D_0  \sum_{i=1}^n Z(t_0, t_i)  \sum_{k =1}^N \frac{\EU_{k}\lambda_k(t_i)\Dr}{D_{tot} - D_k}\left(1+\e_k(t_i)\right)
\ee

We can take the limit of $\Dr \rightarrow 0$ and write the total cost as a sum over clearing members as follows:
\be
\boxed{ C_0(T) =  D_0\sum_{k=1}^N \bar{E}_k \int_{t_0}^{T} \lambda_k(t) Z(0,t) dt  }
\ee
where the exposure is given by:
\be\label{eq-SNexposure}
\boxed{
\bar{E}_k \equiv \frac{w \hp_k}{\alpha-1} \left( \frac{M_k}{M_k + D_k} \right)^\alpha \frac{M_k+D_k}{D_{tot} - D_k} (1+ \e_k)
}
\ee
and we have assumed that $\e_k(t) = \e_k$ is constant in time.


In practice, we are unlikely to have member specific information on the collateral and model parameters and we need to approximate these with average quantities. Let us denote these by  $\hat{p}$, $M$, $G$ and $\lambda$ respectively. We also introduce the duration weighted average intensity:
\be
 \bar{\lambda} \equiv \frac{1}{T- t_0} \int_{t_0}^{T} \lambda(t) Z(0,t) dt
\ee
If, in addition, we assume that $M_k \gg D_k$ and that $\e_k$ is negligible, then we get the following simple expression for the CCP cost:
\be
C_0(T) \approx \frac{w \hat{p}}{\alpha-1} \bar{\lambda} (T-t_0)(M+G).
\ee

We note that the risk is proportional to the wrong-way factor $w$, the breach probability $\hat{p}_k$, the average default intensity, $\bar{\lambda}$, as well as the average collateral. We have an inverse proportionality to the Pareto index $\alpha$. In addition, we see that the risk does not depend on the number of clearing members.

\section{Parameter Estimation}
\subsection{Estimating the Wrong-Way Factor $w$}\label{sec-w}
The wrong-way factor, $w$, determines the ratio of margin levels at the time of member default vs.\ today's margin levels. To estimate the factor we assume that the initial margin and default fund levels are VaR based. This means that the margin level is set by the CCP so that there is a fixed probability of losses exceeding this margin over the liquidation period.
We write the portfolio loss distribution, used \emph{by the CCP} in estimating the margin levels, as a function, $f_k$, of a Gaussian variable $X_t$, i.e. $\DV_k(t) = f_k(X_t)$. The volatility of $X_t$ is the underlying market volatility and denoted by $\sigma(t)$.

Let us denote a collateral level at time $t$ by $C_k(t)$. This could either represent the initial margin or a member's default fund contribution. Our distributional assumption means that for a given clearing member $\CM_k$:
\be
p \equiv P[\DV_k(t-) > C_k(t)] = \Phi\left(-\frac{f^{-1}_k(C_k(t))}{\sigma(t-)}\right)
\ee
where $\Phi(\cdot)$ denotes the standard cumulative normal distribution. It is important to note that when setting the collateral level, the default of the member $k$ has not yet occurred and hence there is no possibility that the member default has had a chance to impact the market volatilities.

The probability of breaching the level $C_k$ is given by $p$ and is fixed for all times by the CCP.
We can use this to back out the collateral level as follows:
\be
C_k(t) = f_k\left( \sigma(t-)g \right)
\ee
where $g = \Phi^{-1}(1-p)$.

The wrong-way factor, $w$, is given by the ratio of collateral levels at time $\tau_k$ vs.\ time $t_0$.
This means that we have:
\be
w = \frac{C_k(\tau_k)}{C_k(t_0)} = \frac{f_k(\sigma(\tau_k-) g)}{f_k(\sigma(t_0)g)}
\ee
 If we assume that $f_k$ is linear then we have simply:
\be
w = \frac{\sigma(\tau_k-)}{\sigma(t_0)}
\ee
This means we can calculate $w$ as a ratio of a stressed volatility over today's volatility. We assume in the following that the stressed volatility does not depend explicitly on time and is given by:
\be
\sigma(\tau_k-) \equiv \sigma^*
\ee

\subsubsection{Historical estimation}\label{sec-estimation-w}
We estimate the wrong-way factor $w$ in four markets: equity, credit, FX, and rates. We do this by estimating a stress volatility of historical 5-day log-returns of the following representative indices:
S\&P 500 (1981 to 2011);
CDX IG (2003 to 2011);
USD/GBP fx rate (1985 to 2011) and
USD 1 year swap rate (1990 to 2011).
This is then compared to today's estimate of the volatility in order to calculate $w$.

Volatilities are estimated for each day of our time-series using a simple exponentially weighted moving average (EWMA) algorithm. This means that the volatility, $\sigma_i$ at date $t_i$ is given by:
\be\label{eq-EWMA_back}
\sigma^2_i = \lambda\sigma_{i-1}^2 + (1-\lambda)r_i^2
\ee
where $r_i$ is the 5-day log return at time $t_i$.
We set the EWMA parameter $\lambda$ to $0.99$. This determines how heavily past observations are weighted relative to more recent ones.

The stress volatility is defined as the 99\% quantile in the historically estimated volatility distribution. A summary  of our estimates for $w$ is given in table~\ref{table-w}. We observe that $w$ ranges between 1.3 and 2.5.

\begin{table}[h]
\begin{center}
\begin{tabular}{|l|l|l|l|l|} \hline
            & S\&P 500  & CDX IG    & USD/GBP   & USD       \\ \hline \hline
current vol (5 day) & 3.3\%     & 6.1\%     & 1.2\%     & 7.9\%     \\ \hline
max vol     & 5.8\%     & 12.5\%    & 3.1\%     & 10.5\%     \\ \hline
99\% vol    & 5.5\%     & 12.3\%    & 2.9\%     & 10.1\%     \\ \hline
wrong-way-factor $w$    & 1.7     & 2.2    & 2.5     & 1.3    \\ \hline
\end{tabular}
\end{center}
\caption{Estimated wrong-way-factor for different markets. To calculate $w$ we take $p=1\%$ and $f(x) \sim \exp(\sigma x)-1$.} \label{table-w}
\end{table}

\subsection{Estimating the breach probability $\hp_k$} \label{sec-p}
The conditional expected tail loss $\EU_k$ depends on the probability $\hp_k$ of losses breaching the initial margin conditional on the member default. To estimate this we can again make use of the fact that for a typical CCP the initial margin level is VaR based.

The distribution used to define the margin level is based on the market volatilities, $\sigma^*$, just prior to the member default.
As in the previous section, we assume that portfolio loss is a function $f_k$ of a Gaussian random variable up to the margin level $M_k^*$.
The probability of breaching the margin is set by the CCP and denoted by $p_M$ (typically this probability is set at $p_M= 1\%$). This means that:
 \be
 p_M \equiv P[\DV_k(\tau_k-) > M^*_k ] = \Phi\left(-\frac{f^{-1}_k(M^*_k)}{\sigma^*}\right)
 \ee

Post-default we expect market stress to increase due to contagion. This means that we expect the loss distribution volatility to increase from the level $\sigma^*$ to an even more stressed level of $\hat{\sigma}_k > \sigma^*$. It is $\hat{\sigma}_k$ that drives the portfolio losses over the unwind period $\Dl$.  We assume that:
\be
\hat{\sigma}_k = \gamma_k \sigma^*
\ee
where $\gamma_k \geq 1$ is the contagion stress factor. In general, this can be specific to the member $k$, which is useful if we believe that some members are more systemically important and hence can cause more market disruption in the case of default.

The probability of losses exceeding $M_k^*$ conditional on default is given by $\hat{p}_k$ and is greater than $p_M$.
We have:
 \begin{eqnarray*}
 \hp_k &\equiv& P[\DV_k > M^*_k | \s] \\
 &=& \Phi\left(-\frac{f^{-1}_k(M^*_k)}{\hat{\sigma}_k}\right) \\
 &=& \Phi\left(\frac{\Phi^{-1}(p_M)}{\gamma_k}\right)
 \end{eqnarray*}
 Note that $\hat{p}_k$ does not depend on $\hat{\sigma}_k$ but only on the stress factor $\gamma_k$ and $p_M$, which is set by the CCP.

 In addition to the contagion effect there are at least two further reasons for why the realised breach probability can be higher than the estimated probability, i.e. why $\hat{p}_k > p_M$.
 \begin{itemize}
 \item Estimating future volatilities based on historical information is  difficult.  The quality of the volatility modeling varies greatly between CCP's. Hence, it is reasonable to expect that estimated probabilities are not always realised.
 \item The liquidation period $\Dl$ that is used by the CCP might be too aggressive. Especially in crisis situations it might take longer to unwind one or more member's portfolios. A longer unwind period results in a higher effective volatility of losses and hence a higher breach probability.
 \end{itemize}

 \subsubsection{Historical estimation}

We estimate $\gamma$ and $\hat{p}$ using the same market data as in section~\ref{sec-estimation-w}. Note that here we only estimate a $\gamma$ factor per market and not per clearing member.
In particular, for each time-series date, $t_i$, we calculate a backward-looking volatility, $\sigma_i$, and a forward-looking volatility, $\sigma_i'$. The backward-looking volatility is calculated using EWMA and defined via equation~\eref{eq-EWMA_back}. To calculate $\sigma_i'$ we use a forward-looking EWMA algorithm, i.e.:
\be\label{eq-EWMA_fwd}
\sigma'^2_i = \lambda'\sigma'^2_{i+1} + (1-\lambda')r_i^2
\ee
In our tests we set $\lambda'=0.97$ as this provided a good estimate of realised exceedence probabilities.

For each day we also compute the ratio $\gamma(t_i) = \sigma'_i / \sigma_i$. Our contagion factor, $\gamma$, is defined as the 99\% quantile in the historical $\gamma(t_i)$ distribution. The results of the estimation are given in table~\ref{table-gamma}. We observe a fairly stable estimation across markets with a range of $\gamma$ between 2.0 and 2.6. This corresponds to stressed breach probabilities, $\hat{p}$ between 12\% and 18\%. We note that the sensitivity of $\hat{p}$ to $\gamma$ is very high.

 \begin{table}[h]
\begin{center}
\begin{tabular}{|l|l|l|l|l|} \hline
            & S\&P 500  & CDX IG    & USD/GBP   & USD       \\ \hline \hline
contagion factor $\gamma$    & 2.1       & 2.0       & 2.3        & 2.6     \\ \hline
breach probability $\hat{p}$ & 14\%     &12\%       &16\%           &18\%   \\ \hline
\end{tabular}
\end{center}
\caption{Estimated contagion factor for different markets.} \label{table-gamma}
\end{table}

\subsection{Estimating the Pareto Index $\alpha$} \label{sec-alpha}
We estimate the Pareto index $\alpha$ by analysing the return time-series in the four markets introduced above. However, here we consider \emph{absolute changes} in the indices over the liquidation period (5 days) as opposed to log-returns. This is because for simple contracts, e.g. forwards or CDS, the loss would be proportional to the absolute change in the underlying.

For each set of  return data, a Pareto distribution is calibrated to match the 99\% quantile of the historical distribution. The index is then optimised to obtain a best least-squares fit of the distribution tails below the 99\% quantile.

\begin{figure}[h]
\begin{center}
\includegraphics[width=7.5cm, height=4.25cm]{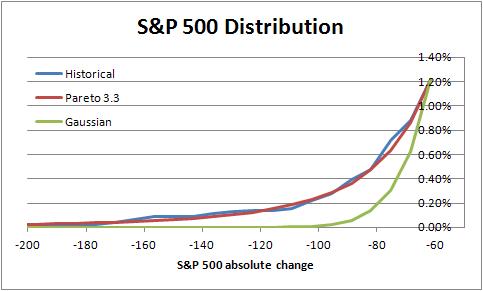}
\includegraphics[width=7.5cm, height=4.25cm]{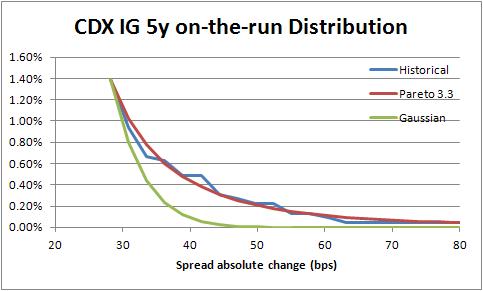}
\includegraphics[width=7.5cm, height=4.25cm]{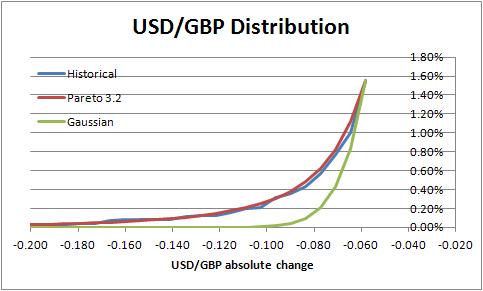}
\includegraphics[width=7.5cm, height=4.25cm]{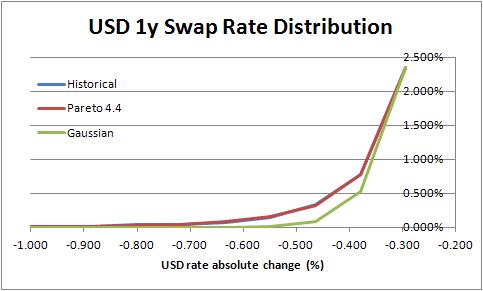}
\caption{Pareto distribution fitted to historical distributions of absolute changes for S\&P 500, CDX, USD/GBP and USD swap rates. Gaussian tail is included for comparison. }\label{fig-alpha}
\end{center}
\end{figure}

Our results can be found in figure~\ref{fig-alpha}.
We observe that the Pareto distribution can provide a good fit in the different markets. We estimate an index of $\alpha = 3.3$ for equity, credit and FX markets. For the USD rates market we observe a higher index of $\alpha = 4.4$. We note that a Gaussian tail corresponds approximately to $\alpha = 7$, so in all cases the observed tails are significantly larger than would be implied by a Gaussian distribution.

\subsection{Estimating the Expected Loss}
We can use our parameter estimates to provide indicative values for the expected allocated loss in the case of a clearing member default. This in turn can be used to calculate a CCP risk charge. We make the following  assumptions in our calculation:
\begin{itemize}
\item Parameters are homogeneous across all clearing members. CDS spreads for the members are are taken to be flat at 200 bps. This implies that the default intensity  is given by $\lambda = 200 \mbox{bps} / (1 - 40\%) \approx 333 \mbox{bps}$ where $40\%$ is a typical recovery rate.
\item Typically, default fund contributions are around $3\%$ to $10\%$ of the posted initial margin. A reasonable correction term is around $\e = 20\%$. Here we assume $(M/(M+G))^\alpha (1+\e) \approx 1$.
\item We have assumed throughout that CCP equity is 0.
\item The liquidation period is set at $\Dl$ = 5 days.
\end{itemize}

In this approximation the 1 year risk charge for a given member as a fraction of the total posted collateral is given by:
\be
\bar{C_0} \equiv \frac{C_0(1)}{M+G} \approx \frac{w \hat{p}\bar{\lambda}}{(\alpha-1)} \equiv LGD_{tot} \bar{\lambda}
\ee
The total loss given default, $LGD_{tot}$, gives the total notional of 1 year CDS protection required to hedge the CCP risk (as a proportion of the total posted collateral).

The results of the calculation are summarised in table~\ref{table-charge}. We observe that $LGD_{tot}$ ranges between 6.9\% and 17.4\%. The corresponding cost of 1y protection ranges between 23 bps and 58 bps. This is comparable to the cost of investment grade super-senior protection.

\begin{table}[h]
\begin{center}
\begin{tabular}{|l|l|l|l|l|} \hline
                            & S\&P 500  & CDX IG    & USD/GBP   & USD       \\ \hline \hline
wrong-way-factor $w$        & 1.7       & 2.2       & 2.5     & 1.3    \\ \hline
breach probability $\hat{p}$ & 14\%     &12\%       &16\%     &18\%   \\ \hline
Pareto index $\alpha$       & 3.3       & 3.3       & 3.3     & 4.4    \\ \hline
Protection notional $LGD_{tot}$ &10.3\% &11.5\%      &17.4\%   & 6.9\%    \\ \hline
1y CCP risk charge $\bar{C_0}$  & 34bps   & 38bps       & 58bps     & 23bps    \\ \hline
\end{tabular}
\end{center}
\caption{Estimated CCP risk over 1 year.} \label{table-charge}
\end{table}

\section{Conclusions and Future Directions}

Membership of a clearinghouse carries risk. In this note we have presented a practical model for quantifying the expected losses that a clearing member can incur on their posted collateral when facing a CCP.

We have shown that the risk is given by a sum of exposures to each of the clearing members. The exposures are driven by the member portfolio loss distribution tails which we have parameterised by a Pareto distribution. This is calibrated to the CCP defined confidence level of losses exceeding the posted initial margin levels.
Key risk factors are the expected level of wrong-way-risk and contagion, which can be estimated by analysing historical market volatilities.

We have shown that for realistic assumptions, the systemic CCP risk a clearing member is exposed to is not negligible. The cost of protection from losses on the posted collateral is comparable to the cost of investment grade super-senior protection.

Several directions for future research present themselves:
\begin{itemize}
\item Alternative parameterisations for the tail behaviour of the portfolio loss distributions can be explored and compared to those of actual portfolios. The applicability of our assumptions
    can be examined in a wider range of markets and for different product types. A more detailed investigation of the risk posed by particular CCPs is also possible.
\item We have not considered how accounting for debt value adjustment (DVA) might impact the CCP risk calculated here. Given that we expect CCP losses to occur in times of systemic stress, it is likely that DVA might provide an offsetting benefit.
\item The assumptions around portfolio rollover and collateral levels through time can be refined.
\item The framework described here can also be used profitably to quantify capital requirements for centrally cleared trades as well as default fund commitments.
\end{itemize}

\end{document}